\documentclass{epl}

\title{What is the value of the superconducting gap \\
of a F/S/F trilayer ?}
\shorttitle{Superconducting gap of a F/S/F trilayer}
\author{R. M\'elin\inst{1}\thanks{melin@grenoble.cnrs.fr},
D. Feinberg\inst{2}}
\institute{
\inst{1} Centre de Recherches sur les Tr\`es Basses
Temp\'eratures (CRTBT\thanks{U.P.R. 5001 du CNRS, Laboratoire
conventionn\'e avec l'Universit\'e Joseph Fourier}),\\ CNRS, BP 166,
38042 Grenoble Cedex 9, France\\
\inst{2} Laboratoire d'Etude des Propri\'et\'es Electroniques
des Solides (LEPES\thanks{U.P.R. 0011 du CNRS, Laboratoire
conventionn\'e avec l'Universit\'e Joseph Fourier}),\\ CNRS, BP 166,
38042 Grenoble Cedex 9, France}


\begin{document}
\maketitle

\begin{abstract}
Based on the model of F/S/F trilayer with atomic
thickness~\cite{atomic}
we discuss the relative roles of pair-breaking and proximity effects, 
as a function of the exchange field, of
disorder and of a finite
thickness in the superconducting layer. 
The exchange field can be small (weak
ferromagnets) or large (strong ferromagnets) compared
to the superconducting gap. 
With weak ferromagnets 
we show the existence of a reentrant superconducting gap
for the F/S/F trilayer with atomic thickness
in the parallel alignment (equivalent to the F/S bilayer).
Qualitatively small disorder is equivalent to reducing the value of
the hopping parameters.
In the presence of a finite thickness
in the superconducting layer
the superconducting gap in the antiparallel alignment is larger
than in the parallel alignment, meaning that pair breaking
dominates over the proximity effect. 
\end{abstract}

\section{Introduction}

Many interesting phenomena take place in mesoscopic devices
when superconductors
are connected to ferromagnets. For example
the superconducting pair amplitude induced in a ferromagnet
oscillates in
space~\cite{Fulde-Ferrel,Larkin,Clogston,Demler} and these oscillations
can give rise to the $\pi$-coupling. 
The $\pi$-coupling
manifests itself in superconductor / ferromagnet
(S/F) multilayers
as oscillations of the superconducting
critical temperature with the thickness of the 
ferromagnetic layers~\cite{Buzdin2,exp1,exp5}.
S/F/S $\pi$-junctions have been recently probed with various
experimental techniques~\cite{Ryazanov,Kontos,Gandit}.

The proximity effect in F/S/F trilayers has recently focused a renewed
interest.
It was established long ago that with insulating
ferromagnets the superconducting transition temperature
is larger in the antiparallel
alignment because of the
exchange field induced in the S layer~\cite{deGennes} that
tends to dissociate Cooper pairs. Following this
theoretical prediction two experiments were performed, one
with metallic ferromagnets~\cite{Deutscher} and the other with
insulating ferromagnets~\cite{Hauser} and it was shown in
both cases that the critical temperature was larger in the
antiparallel spin orientation.
On the other hand with metallic
ferromagnets the theoretical description should also include
the proximity effect, namely the possibility that
Cooper pairs from the superconductor can delocalize in the 
ferromagnetic electrodes. The proximity effect in
F/S/F trilayers is unusual because it may involve
spatially separated superconducting correlations if the S layer is 
thin enough~\cite{Melin,atomic}.
The basic physics can be understood from considering a model
of half-metal ferromagnets: in
the parallel spin orientation Cooper pairs from the
superconductor cannot be transfered in the ferromagnetic
electrodes. As a consequence the zero temperature
superconducting gap is not affected by the coupling to the
ferromagnetic electrodes~\cite{atomic}.
In the antiparallel spin orientation Cooper pairs from the
superconductor can delocalize in the ferromagnetic electrodes:
the spin-up electron can tunnel
in the spin-up electrode and the spin-down electron can tunnel
in the spin-down electrode~\cite{df}. 
The superconducting gap is reduced by this proximity effect 
associated to spatially separated superconducting
correlations~\cite{Melin} and
the zero-temperature superconducting gap is
larger in the parallel alignment
($\Delta_{\rm P}>\Delta_{\rm AP}$)~\cite{atomic,Melin}.
On the other hand it was shown
that the superconducting
transition temperature with metallic ferromagnets
is, like with insulators, larger in the antiparallel
alignment~\cite{Baladie} ($T_c^{\rm P}<T_c^{\rm AP}$)
which has been probed in recent
experiments~\cite{Gu}. 
In a recent work Buzdin and Daumens~\cite{atomic} proposed to reconcile 
these apparently contradictory results for the gap at $T=0$ and for $T_c$, by 
calculating
the critical temperature of a F/S/F trilayer with atomic thickness
within the Stoner model, and the zero temperature gap
within a model
of half-metal ferromagnet. They indeed found that $T_c^{\rm P}<T_c^{\rm AP}$
but $\Delta_{\rm P}>\Delta_{\rm AP}$. 
This suggests that the proximity effect plays a dominant role in the
determination of the zero temperature superconducting gap and that
pair breaking effects play a dominant role in the determination
of the critical temperature. 
The goal of this Letter is to determine whether this picture is robust,
including realistic ingredients such as
a strong or weak exchange field in the ferromagnets,
disorder or a finite thickness in the superconductor.

\section{The model}

Let us start with the F/S/F trilayer with atomic
thickness~\cite{atomic}. The superconducting layer is described
by the BCS Hamiltonian
\begin{equation}
{\cal H}_{\rm BCS} = \sum_{\langle \alpha , \beta
\rangle , \sigma} - t \left(
c_{\alpha,\sigma}^+ c_{\beta,\sigma}
+c_{\beta,\sigma}^+ c_{\alpha,\sigma} \right)
+ \sum_\alpha \left( \Delta_\alpha
c_{\alpha,\uparrow}^+ c_{\alpha,\downarrow}^+
+ \Delta_\alpha^* c_{\alpha,\downarrow}
c_{\alpha,\uparrow} \right)
,
\end{equation}
where $\alpha$ and $\beta$ correspond to neighboring sites on a
square lattice.
The ferromagnetic electrodes are described by the Stoner
model
\begin{equation}
\label{eq:Stoner}
{\cal H}_{\rm Stoner} = \sum_{\langle \alpha,\beta
\rangle,\sigma} -t \left( c_{\alpha,\sigma}^+
c_{\beta,\sigma} + c_{\beta,\sigma}^+ 
c_{\alpha,\sigma} \right)
-h_{\rm ex} \sum_i 
\left( c_{\alpha,\uparrow}^+ c_{\alpha,\uparrow}
-c_{\alpha,\downarrow}^+ c_{\alpha,\downarrow} \right)
,
\end{equation}
where $h_{\rm ex}$ is the exchange field.
The ferromagnetic layers a and b are connected
to the superconducting layer by the Hamiltonian
\begin{equation}
\label{eq:tunnel-ab}
{\cal W}_{a (b)} = \sum_{\alpha,\sigma} - t_{a (b)} \left(
c_{\alpha,\sigma,a (b)}^+ c_{\alpha,\sigma,S} +
c_{\alpha,\sigma,S}^+ c_{\alpha,\sigma,a (b)} \right) 
.
\end{equation}
Strickly speaking the individual layers of the F/S/F
trilayer with atomic thickness might
be unstable against charge or spin density wave ordering. We view the
F/S/F trilayer with atomic thickness as a toy model for
more realistic models involving a finite thickness in the F and S
layers so that we can safely neglect these instabilities.

The Green's functions of the superconductor are
given by~\cite{Abrikosov}
\begin{eqnarray}
\label{eq:galpha11}
g_{\alpha,\alpha}^{1,1}(\xi,\omega)&=&
\frac{u_p^2}{\omega-\epsilon(p)+i\delta}
+\frac{v_p^2}{\omega+\epsilon(p)-i\delta}\\
\label{eq:falpha12}
f_{\alpha,\alpha}^{1,2}(\xi,\omega)&=&
-\frac{\Delta}
{\left[\omega-\epsilon(p)+i\delta\right]
\left[\omega+\epsilon(p)-i\delta\right]}
,
\end{eqnarray}
and similar expressions are obtained for $g_{\alpha,\alpha}^{2,2}$
and $f_{\alpha,\alpha}^{2,1}$. The labels ``1'' and ``2'' in
(\ref{eq:galpha11}) and (\ref{eq:falpha12}) correspond to the
Nambu indexes. $\epsilon(p)$ corresponds to the quasiparticle
energy $\epsilon(p)=\sqrt{\Delta^2+\xi^2}$, with
$\xi =\hbar^2 p^2/(2m)-\epsilon_F$ the kinetic energy determined
with respect to the Fermi level. The coherence factors are given
by $u_p^2=(1+\xi/\epsilon(p))/2$ and
$v_p^2=(1-\xi/\epsilon(p))/2$.
The ``11'' Green's function of a spin-up ferromagnetic electrode is
given by
$g_{a,a}^{1,1} = 1/[\omega-\xi-h_{\rm ex}
+i \delta \mbox{ sgn}(\xi+h_{\rm ex})]$
,
and a similar expression is obtained for the ``22'' component.

\section{Perturbative expansions}

The fully dressed Green's function 
corresponding to the F/S/F trilayer with atomic thickness is 
determined through the Dyson equation
$
\hat{G}_{\alpha,\alpha} = \hat{g}_{\alpha,\alpha}
+\hat{g}_{\alpha,\alpha} \hat{t}_{\alpha,a}
\hat{g}_{a,a} \hat{t}_{a,\alpha} \hat{G}_{\alpha,\alpha} 
+ \hat{g}_{\alpha,\alpha} \hat{t}_{\alpha,b}
\hat{g}_{b,b} \hat{t}_{b,\alpha} \hat{G}_{\alpha,\alpha}
$.
The self-consistency relation takes the form~\cite{Abrikosov}
\begin{equation}
\Delta=\lambda \int \frac{d\omega}{2\pi}
\frac{d^2{\bf k}}{(2\pi)^2} G_{\alpha,\alpha}^{1,2}({\bf k},\omega)
.
\end{equation}
To order $t^2$ and for strong ferromagnets 
($\Delta_0 \ll t \ll h_{\rm ex}$) we obtain:
\begin{equation}
\ln{\left(\frac{\Delta}{\Delta_0}\right)}
=-2 \frac{t_a^2+t_b^2}{h_{\rm ex}^2}
\left[\ln{\left(\frac{h_{\rm ex}}{\Delta_0}\right)}
-\frac{1}{2} \right]
,
\end{equation}
where $\Delta_0$ is the superconducting gap of the isolated
superconductor.
In the case of weak ferromagnets
($t \ll h_{\rm ex} < \Delta_0$) we obtain
\begin{equation}
\label{eq:pert-weak}
\ln{\left(\frac{\Delta}{\Delta_0}\right)}
=-\frac{1}{2} \frac{(t_a^2+t_b^2) h_{\rm ex}^2}{\Delta_0^4}
.
\end{equation}
We deduce that in both cases the superconducting gap at order $t^2$
is reduced by the proximity with the ferromagnetic layers,
independently of the 
relative spin polarizations of the F layers.

In the case of strong ferromagnets a perturbation theory
to fourth order in $t$ leads to
\begin{eqnarray}
\ln{\left(\frac{\Delta_{\rm P}}{\Delta_{\rm AP}}\right)}
=2 \frac{t_a^2 t_b^2}{h_{\rm ex}^4}
\left[ 7 \ln{
\left(\frac{h_{\rm ex}}{\Delta_0}\right)}
-4- \ln{\left(\frac{\Delta_0}{\eta}\right)} \right]
,
\end{eqnarray}
where we introduced a small cut-off $\eta$ to regularize the
logarithmic divergence of the integral over $\xi$.
The (positive) difference between the superconducting gap in the
parallel and antiparallel alignments is of order
$(t/h_{\rm ex})^4$ and is thus very small compared to $\Delta_0$.

For weak ferromagnets a perturbation theory to
order $t^4$ leads to
\begin{equation}
\label{eq:expan-weak}
\ln{\left(\frac{\Delta_{\rm P}}{\Delta_{\rm AP}}\right)}
= 2 \frac{t_a^2 t_b^2}{\Delta_0^4}\left[
\frac{3}{2} + \ln{\left(\frac{4 h_{\rm ex} \eta}{\Delta_0^2}
\right)} \right]
+ 2 \frac{t_a^2 t_b^2 h_{\rm ex}^2}{\Delta_0^6}
\left[ -\frac{19}{6} + 2 \ln{\left(\frac{2 h_{\rm ex}^2}
{\Delta_0 \eta} \right)} \right] + ...
\end{equation}
In this case
there are two small parameters in
perturbation theory: 
$t^2/\Delta_0^2$ and $h_{\rm ex}^2/\Delta_0^2$.
Perturbation theory is well suited for understanding
the small parameters in the problem but is not
quantitatively reliable because of
logarithmic divergences in the limit $\eta \rightarrow 0$.

\begin{figure}
\onefigure[width=12cm]{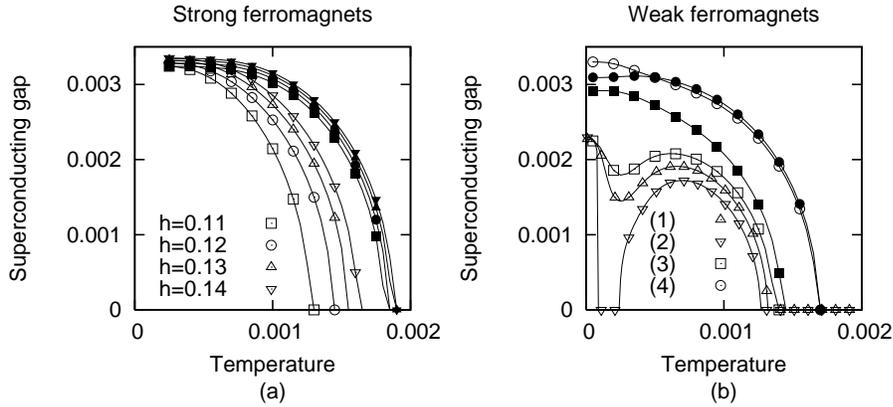}
\caption{Variation of the superconducting gap $\Delta(T)$
as a function of temperature $T$ for the F/S/F trilayer
with atomic thickness (a) for strong ferromagnets
and (b) for weak ferromagnets and with the parameters
$\hbar=1$, $m=2$, $k_F=1$, $\lambda=0.32$.
The Debye frequency is set to $\omega_D=\hbar^2 k_F^2/(2m)$.
For both figures we use $t=t_a=t_b$.
For strong ferromagnets we used
$t=0.01$. For weak ferromagnets we used $h=0.003$ and
$t=0.00235$ (1), $t=0.00230$ (2),
$t=0.00225$ (3) and $t=0.00175$ (4).
The open symbols
correspond to the parallel alignment and the filled
symbols correspond to the antiparallel alignment.
}
\label{fig:Delta}
\end{figure}

\section{Numerical simulations of the F/S/F trilayer
with atomic thickness}

Non-perturbative solutions valid for arbitrary interface transparencies
and for finite temperatures can be implemented numerically. 
For a F/S/F trilayer with atomic thickness we can solve explicitly
the Dyson equation and obtain the expression of the
dressed propagator $G_{\alpha,\alpha}^{1,2}$:
$G_{\alpha,\alpha}^{1,2}=f_{\alpha,\alpha}^{1,2}/{\cal D}$, with
\begin{equation}
{\cal D} = 1 - g_{\alpha,\alpha}^{1,1} 
\Sigma_t^{1,1}-g_{\alpha,\alpha}^{2,2} \Sigma_t^{2,2}
+ \left[ g_{\alpha,\alpha}^{1,1} g_{\alpha,\alpha}^{2,2}
-f_{\alpha,\alpha}^2 \right] \Sigma_t^{1,1} \Sigma_t^{2,2}
\label{eq:D}
,
\end{equation}
where the self-energy $\Sigma_t$ is diagonal in Nambu space:
$\Sigma_t^{\tau_1,\tau_2}=(t_a^2 g_{a,a}^{\tau_1,\tau_1}
+t_b^2 g_{b,b}^{\tau_1,\tau_1})\delta_{\tau_1,\tau_2}$,
where $\tau=1,2$ is a Nambu index.

The temperature dependence of
the superconducting gap in the case of strong ferromagnets is shown
on Fig.~\ref{fig:Delta}-(a).
We obtain $\Delta_{\rm P}(0) \simeq
\Delta_{\rm AP}(0)$ but $T_c^{\rm AP} > T_c^{\rm P}$.
Because electrons in the S layer can make excursions in the F layers
an exchange field of order
$(t_a^2+t_b^2)/h_{\rm ex}$
is induced in the superconducting layer in the parallel
alignment~\cite{deGennes,Volkov}. This pair-breaking mechanism 
is absent in the antiparallel case. The order of magnitude of the
exchange field
is obtained by evaluating the density of spin-up
and spin-down electrons in the superconductor.
Superconductivity breaks down at zero temperature if the 
exchange field in the F layer becomes smaller than
a critical exchange field $h_{\rm ex}^{(0)}$ of order
$h_{\rm ex}^{(0)} \simeq (t_a^2+t_b^2)/\Delta_0(0)$.
For the parameters corresponding to Fig.~\ref{fig:Delta} we
obtain from the numerical simulation
$h_{\rm ex}^{(0)} = 0.092 \pm 0.002$ which is of order
$2 t^2 / \Delta_0$.
The F/S/F trilayer 
in the parallel alignment with the parameters on
Fig.~\ref{fig:Delta}-(a) is close to the
breakdown of superconductivity and
this is why we obtain large values of $T_c^{\rm P}-T_c^{\rm AP}$.
This regime may be of interest from the point of view of realizing
a superconducting spin valve.
Yet, this would require a fine tuning
of some parameters such as the interface transparencies.

With weak ferromagnets (see Fig.~\ref{fig:Delta}-(b))
the most interesting case corresponds to
$h_{\rm ex}$ slightly smaller than $\Delta_0(0)$.
We obtain  $\Delta_{\rm P}(0)>\Delta_{\rm AP}(0)$ for small values of
$t=t_a=t_b$ and a crossing between $\Delta_{\rm P}(T)$
and $\Delta_{\rm AP}(T)$.
For larger values of $t$ we obtain
$\Delta_{\rm P}(T)<\Delta_{\rm AP}(T)$ for all values of
the temperature. If the interface transparencies are
sufficiently large the gap $\Delta_{\rm P}(T)$
in the parallel alignment
is minimal at $T=T^*$ and reenters at a lower temperature.
This behavior is apparently due to a cross-over between
two regimes: $\Delta_0(T)>h_{\rm ex}$ for $T<T^*$
(weak ferromagnets) where the small parameter
in perturbation theory is $h_{\rm ex}/\Delta_0(T)$
and $\Delta_0(T)<h_{\rm ex}$ for $T>T^*$ (strong ferromagnets),
where the small parameter in perturbation theory is
$\Delta_0(T)/h_{\rm ex}$. Given the perturbative expansion like
(\ref{eq:pert-weak}) we see that the pair-breaking effect of magnetism
onto superconductivity is maximal  for 
$h_{\rm ex} \simeq \Delta_0(T)$. We also verified that for
smaller values of $h_{\rm ex}$ we obtain larger values
of $T^*$ which is in agreement with this scenario.

\begin{figure}
\onefigure[width=12cm]{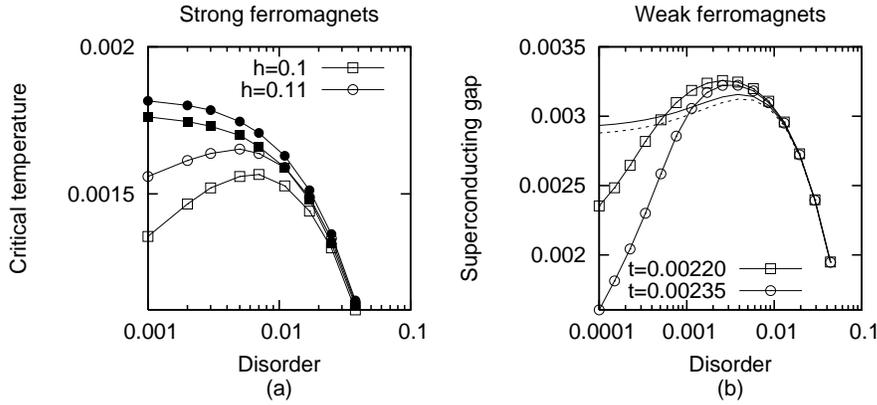}
\caption{ (a) Variation of the critical temperature
as a function of disorder in the S layer for strong
ferromagnets.
(b) Variation of the superconducting gap as a function
of disorder in the S layer for weak ferromagnets.
Both cases correspond to the F/S/F trilayer with
atomic thickness.
We used the parameters
$\hbar=1$, $m=2$, $k_F=1$, $\lambda=0.32$.
The Debye frequency is set to $\omega_D=\hbar^2 k_F^2/(2m)$.
For strong ferromagnets we used
$t=t_a=t_b=0.01$. For weak ferromagnets we used $h=0.003$ and
the temperature is set to $T=0.00025$.
In (a) the open symbols
correspond to the parallel alignment and the filled
symbols correspond to the antiparallel alignment.
In (b) the open symbols correspond to the parallel
alignment. The solid line corresponds to the antiparallel
alignment with $t=0.0220$ and the dashed line corresponds
to the antiparallel alignment with $t=0.0235$.
Disorder is measured by $\Sigma^{1,1}_\alpha(\omega)$
for $\omega \gg \Delta$.
}
\label{fig:dis}
\end{figure}

\section{Effect of disorder}
To include disorder in the superconducting 
and ferromagnetic layers we cannot use Usadel equations
since we want to describe spatially separated superconducting
correlations. Instead we use a perturbative treatment of
disorder based on Ref.~\cite{Abrikosov}. 
It is well known that the superconducting gap of an isolated
superconductor
is not affected by non magnetic impurities~\cite{Abrikosov}
and we could verify this in our simulations.
We show here that
this is not the case in the F/S/F trilayer and that
introducing  non magnetic impurities has the same effect as reducing 
the value of the hopping parameters.
The self-energy associated to disorder
in the superconductor takes the form
$\Sigma_\alpha^{1,1} = n_\alpha u_\alpha^2 g_{\alpha,\alpha}^{1,1}(0)$,
where $n_\alpha$ denotes the concentration of impurities,
$u_\alpha$ is the scattering potential,
supposed to be isotropic, and $g_{\alpha,\alpha}^{1,1}(0)$
is equal to the propagator of the superconductor evaluated at
$R=0$. The '11' component of the total self-energy is given by
$
\Sigma_\alpha^{1,1,{\rm tot}} = \Sigma_\alpha^{1,1}
+t_a^2 g_{a,a}^{1,1} + t_b^2 g_{b,b}^{1,1}
+t_a^2 \left(g_{a,a}^{1,1}\right)^2 \Sigma_a^{1,1}
+t_b^2 \left(g_{b,b}^{1,1}\right)^2 \Sigma_b^{1,1}
$, and a similar expressions are obtained for
$\Sigma_\alpha^{2,2, {\rm tot}}$.
The '12' component of the total self-energy is given by
$\Sigma_\alpha^{1,2,{\rm tot}}=\Sigma_\alpha^{1,2}$, with
$\Sigma_\alpha^{1,2} = n_\alpha u_\alpha^2 g_{\alpha,\alpha}^{1,2}(0)$.
$\Sigma_a$ and
$\Sigma_b$ correspond to the self-energies due to disorder in the
ferromagnetic electrodes.
The expression of the '12'
component of the averaged dressed propagator is given by
\begin{equation}
\overline{G}_{\alpha,\alpha}^{1,2}=
\frac{1}{\cal D} \left\{
f_{\alpha,\alpha}^{1,2}-
\left[ g_{\alpha,\alpha}^{1,1} g_{\alpha,\alpha}^{2,2}
- \left( f_{\alpha,\alpha}^{1,2}\right)^2 \right] \Sigma_\alpha^{1,2}
\right\}
,
\end{equation}
where ${\cal D}$ is given by (\ref{eq:D}) with
$\Sigma_t$ being replaced by $\Sigma_\alpha^{\rm tot}$.
We implement the simplest approximation: we neglect
the corrections to the tunnel vertex in which a
quasiparticle from S scatters on an impurity, tunnels in a
ferromagnetic electrode, tunnels back in the superconductor
and scatters on the same impurity.
In the case of strong
ferromagnets we find that
that a small disorder tends to increase the value of $T_c^{\rm P}$, which
means
a reduction of pair-breaking effects (analogous to a reduction of $t$).
This is absent
in the antiparallel case where disorder reduces the value of 
$T_c^{\rm AP}$ (see Fig.~\ref{fig:dis}-(a)).
For larger values of disorder we have $T_c^{\rm AP}
\simeq T_c^{\rm P}$.
For weak ferromagnets we carried out a simulation
at a temperature close to the minimum on Fig.~\ref{fig:Delta}-(b).
We find that a small disorder tends to increase
the value of the superconducting gap in the parallel and
antiparallel alignments (see Fig.~\ref{fig:dis}-(b)).
The increase of the superconducting gap is larger in the
parallel alignment so that there exists a value of
disorder above which the superconducting gap in the parallel
alignment is larger than the superconducting gap in the
antiparallel alignment (see Fig.~\ref{fig:dis}-(b)).
For larger values of disorder we have
$\Delta_{\rm P}(T)\simeq\Delta_{\rm AP}(T)$.

\begin{figure}
\onefigure[width=12cm]{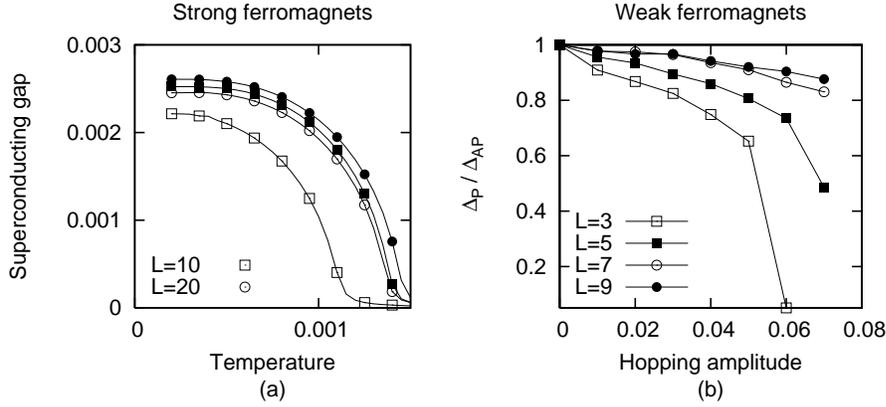}
\caption{ (a) Variation of the superconducting gap
as a function of temperature for strong ferromagnets
for two values of the thickness of the S layer.
(b) Variation of $\Delta_{\rm P}(T)/\Delta_{\rm AP}(T)$
as a function
of the hopping amplitude $t=t_a=t_b$ for weak ferromagnets.
We used the parameters
$\hbar=1$, $m=2$, $k_F=1$, $\lambda=0.32$, $t'=0.1$.
The Debye frequency is set to $\omega_D=\hbar^2 k_F^2/(2m)$.
For strong ferromagnets we used
$t=t_a=t_b=0.02$ and $h_{\rm ex}=0.1$.
For weak ferromagnets we used $h_{\rm ex}=0.003$ and
the temperature is set to $T=0.0004$.
In (a) the open symbols
correspond to the parallel alignment and the filled
symbols correspond to the antiparallel alignment.
In both cases we calculated the superconducting in the
middle of the S layer.
}
\label{fig:thick}
\end{figure}

\section{Effect of a finite thickness in the S layer}

The model of trilayer can be generalized to incorporate
a finite thickness in the superconductor.
This is done by inverting numerically the Dyson matrix 
associated to a chain of
$L$ coupled superconducting layers
$\alpha_0$ , ..., $\alpha_{L-1}$. The layer $\alpha_i$
and $\alpha_{i+1}$ are coupled by a tunnel amplitude $t'$.
The layer $\alpha_0$ ($\alpha_{L-1}$)
is coupled to the left (right) to a ferromagnetic
layer $a$ ($b$) by a tunnel amplitude $t$. 

We have shown on Fig.~\ref{fig:thick}-(a) the variations of
the superconducting gap for different values of the thickness
of the superconducting layer. 
We have $\Delta_{\rm P}(T) < \Delta_{\rm AP}(T)$ meaning that
pair breaking dominates at any temperature for sufficiently
large thicknesses. In the case of weak ferromagnets with
$h_{\rm ex}$ of order $\Delta_0$
we find also  $\Delta_{\rm P}(T) < \Delta_{\rm AP}(T)$
for $t \gg h_{\rm ex}$ 
(see Fig.~\ref{fig:thick}-(b)).
%
We calculated the temperature dependence
of the superconducting gap
for weak ferromagnets with $L_S=2$ layers in the superconductor
and $L_F=2$ layers in the ferromagnets and found a reentrant behavior.
By contrast we found no reentrant behavior for $(L_S,L_F)=(2,1)$.
A systematic study of reentrance as a function of 
$L_S$ and $L_F$ will be presented elsewhere.

\section{Conclusions}
To summarize we found that two regimes of parameters
are relevant from the point of view of possible
experiments: 
strong ferromagnets with tunnel
interfaces such that $\Delta_0 \ll t \ll h_{\rm ex}$
and weak ferromagnets with tunnel interfaces such that
$t \ll h_{\rm ex} < \Delta_0$ for F/S/F trilayers with
atomic thickness, or $h_{\rm ex}\simeq \Delta_0 \ll t$
with a finite thickness in the S layer.
Quantitative predictions can be obtained from
the numerical determination
of the superconducting gap at finite temperature.
For the F/S/F trilayer with atomic thickness
and with strong ferromagnets
we obtain $\Delta_{\rm P}(0) \simeq \Delta_{\rm AP}(0)$
but $T_c^{\rm P}>T_c^{\rm AP}$. With weak ferromagnets
we obtain a reentrant superconducting gap in the F/S
bilayer with atomic thickness.
We discussed the effect of disorder within
the simplest approximation in which we neglect vertex corrections.
In the ferromagnetic alignment and close to the breakdown of
superconductivity we find that disorder increases $T_c^{\rm P}$ in
the case of strong ferromagnets, and $\Delta_{\rm P}$ in the
case of weak ferromagnets. In the presence of a finite thickness
in the superconducting layer pair breaking effects
dominate and we find $\Delta_{\rm P}(T)
<\Delta_{\rm AP}(T)$.

\stars

The authors acknowledge fruitful discussions with A. Buzdin.


\begin{thebibliography}{99}
\bibitem{atomic} A. Buzdin and M. Daumens, cond-mat/0305320,
Europhysics Letters to appear

\bibitem{Fulde-Ferrel} P. Fulde and A. Ferrel,
Phys. Rev. {\bf 135}, A550 (1964).

\bibitem{Larkin} A. Larkin and Y. Ovchinnikov,
Sov. Phys. JETP {\bf 20}, 762 (1965).

\bibitem{Clogston} M.A. Clogston,
Phys. Rev. Lett. {\bf 9}, 266 (1962).

\bibitem{Demler} E.A. Demler,
G.B. Arnold and M.R. Beasley, Phys. Rev. B
{\bf 55}, 15174 (1997).

\bibitem{Buzdin2}
A.I. Buzdin and M. Yu. Kupriyanov,
JETP Lett. {\bf 52}, 487 (1990);

A.I. Buzdin, M. Yu.
Kupriyanov and B. Vujicic,
Physica C {\bf 185 - 189}, 2025 (1991).

\bibitem{exp1} J.S. Jiang, D. Davidovi\'c,
D.H. Reich, and C.L. Chien,
Phys. Rev. Lett. {\bf 74}, 314 (1995).

\bibitem{exp5} Th. Muhge, N.N. Garif'yanov,
Yu. V. Goryunov, G.G. Khaliullin, L.R. Tagirov,
K. Westerholt, I.A. Garifullin, and
H. Zabel, Phys. Rev. Lett. {\bf 77}, 1857 (1996).

\bibitem{Ryazanov} V.V. Ryazanov, V.A. Oboznov,
A. Yu. Rusanov, A.V. Veretennikov,
A.A. Golubov, J. Aarts, Phys. Rev. Lett.
{\bf 86}, 2427 (2001).

\bibitem{Kontos} T. Kontos, M. Aprili,
J. Lesueur, and X. Grison,
Phys. Rev. Lett. {\bf 86}, 304 (2001).

\bibitem{Gandit} W. Guichard, M. Aprili, O. Bourgeois, T. Kontos,
J. Lesueur and P. Gandit, Phys. Rev. Lett. {\bf 90}, 167001 (2003).

\bibitem{deGennes} P.G. de Gennes, Phys.
Letters {\bf 23}, 10 (1966).

\bibitem{Deutscher} G. Deutscher and F. Meunier,
Phys. Rev. Lett. {\bf 22}, 395 (1969).

\bibitem{Hauser} J.J. Hauser, Phys. Rev. Lett.
{\bf 23}, 374 (1969).

\bibitem{Melin} R. M\'elin, J. Phys.: Condens. Matter
{\bf 13}, 6445 (2001);

V. Apinyan and R. M\'elin,
Eur. Phys. J. B {\bf 25}, 373 (2002);

H. Jirari,
R. M\'elin and N. Stefanakis, Eur. Phys. J. B
{\bf 31}, 125 (2003).

\bibitem{df} G. Deutcher and D. Feinberg, Appl. Phys. Lett. 
{\bf 76}, 487 (2000).

\bibitem{Baladie} I. Baladi\'e and A. Buzdin,
Phys. Rev. B {\bf 67}, 014523 (2003).

I. Baladi\'e, A. Buzdin,
N. Ryzhanova, and A. Vedyayev,
Phys. Rev. B {\bf 63}, 054518 (2001).

A. Buzdin, A.V. Vedyayev, and N. Ryzhanova,
Europhys. Lett. {\bf 48}, 686 (1999).

\bibitem{Gu} 
J. Y. Gu, C.-Y. You, J. S. Jiang, J. Pearson,
Ya. B. Bazaliy, and S. D. Bader,
Phys. Rev. Lett. {\bf 89}, 267001 (2002) 

\bibitem{Abrikosov} A.A. Abrikosov, L.P. Gorkov and I.E.
Dzyaloshinski, {\sl Methods of quantum field theory in
statistical physics}, Dover Publications, Inc, New York (1963).

\bibitem{Volkov} F.S. Bergeret, A.F. Volkov and K.B. Efetov,
cond-mat/0307468.

\end{thebibliography}
\end{document}